\documentclass{amsart}

\usepackage{mathptmx}       
\usepackage{helvet}         
\usepackage{courier}        
\usepackage{type1cm}        
\usepackage{graphicx}        
\usepackage{multicol}        
\usepackage[bottom]{footmisc}

\usepackage{url}

\usepackage{pgfplots}
\usepackage{pgfplotstable}
\usepgfplotslibrary{colormaps}
\pgfplotsset{compat=1.15,samples=1000}

\newcommand{\tfidf}{tf-idf}

\newcommand{\summaxprob}{Restricted Maximum Sum Probability}
\newcommand{\summaxprobAc}{RMS}
\newcommand{\maxsumprob}{Maximum Sum Probability}
\newcommand{\maxsumprobAc}{MS}
\newcommand{\maxweightedavgprob}{Maximum Weighted Average Probability}
\newcommand{\maxweightedavgprobAc}{MWA}

\newcommand{\aggregatemethod}{Aggr. method}

\newcommand{\precisionweightedAc}{Prec.}

\newcommand{\recallweightedAc}{Rec.}

\newcommand{\fscoreweightedAc}{F1}
\newcommand{\expiai}{Seg}
\newcommand{\expiaidoc}{Doc}

\newcommand{\expib}{Seg}
\newcommand{\expibdoc}{Doc}

\newcommand{\SVAE}{SVAE}
\newcommand{\SVAEonelayer}{SVAE}
\newcommand{\SVAEtwolayer}{SVAE}

\newcommand{\neuralnet}{NN}

\newcommand{\neuralnetonelayerCPU}{NN}

\newcommand{\randomforest }{RF}
\newcommand{\supportvectorclassifier}{SVM}
\newcommand{\logisticregression}{LR}

\newcommand{\clfactivation}{Activation}
\newcommand{\clfencoderdim}{Encoder dim}
\newcommand{\clflatentdim}{Rel. latent dim}
\newcommand{\clfniternochange}{Iter no change}
\newcommand{\clfsampleweightclf}{Weight CLF}
\newcommand{\clfsampleweightvae}{Weight VAE}
\newcommand{\clftol}{Tolerance}
\newcommand{\clfratioi}{Ratio layer 2}

\newcommand{\clfC}{Regularization}

\newcommand{\clfpenalty}{Penalty}

\newcommand{\clflayeri}{\# layer size 1}
\newcommand{\clflayerii}{\# layer size 2}
\newcommand{\clflayeriii}{\# layer size 3}
\newcommand{\clflearningrateinit}{Init. learning rate}

\newcommand{\clfmaxdepth}{Max tree depth}
\newcommand{\clfnestimators}{\# of trees}

\newcommand{\clfgamma}{Gamma}
\newcommand{\clfkernel}{Kernel}

\title{Study on Text Classification for Public Administration}
\author{Stefanie Schwaar}
\address{Fraunhofer ITWM, Fraunhofer Platz 1, Kaiserslautern}
\address{\emph{and} University of Applied Sciences (HTW Berlin), Berlin, Germany}
\email{stefanie.schwaar@htw-berlin.de}

\author{Franziska Diez}
\address{Fraunhofer ITWM, Fraunhofer Platz 1, Kaiserslautern}
\email{franziska.diez@itwm.fraunhofer.de}

\author{Michael Trebing}
\address{Fraunhofer ITWM, Fraunhofer Platz 1, Kaiserslautern}
\email{michael.trebing@itwm.fraunhofer.de}
\author{Nils Witznick}
\address{Fraunhofer ITWM, Fraunhofer Platz 1, Kaiserslautern}

\begin{document}

\begin{abstract}In German public administration, there are 45 different offices to which incoming messages need to be distributed. Since
these messages are often unstructured, the system has to be based at least partly on message content. For public service
no data are given so far and no pretrained model is available. The data we used are conducted by Governikus KG and are
of highly different length. To handle those data with standard methods different approaches are known, like normalization
or segmentation. However, text classification is highly dependent on the data structure, a study for public administration
data is missing at the moment. We conducted such a study analyzing different techniques of classification based on segments,
normalization and feature selection. Thereby, we used different methods, this means neural nets, random forest, logistic regression, SVM classifier and SVAE.
The comparison shows for the given public service data a classification accuracy of above 80\% can be reached based on
cross validation. We further show that normalization is preferable, while the difference to the segmentation approach depends mainly on the choice of algorithm.
\end{abstract}
\maketitle

\section{Introduction}
\label{sec:introduction}

This study focuses on using machine learning to automatically categorize digital documents, which has become increasingly
important as more and more processes in modern societies are digitized.
With the rise of electronic communication, official documents are now primarily transmitted digitally rather than through
traditional methods like mail or fax.
In administrative environments, this shift allows for the efficient sorting of documents to the department in charge of the
topics they deal with.
This calls for statistical models that accurately assign incoming documents to specific offices.
In the machine learning world this is known as text classification.

The research area of text classification is a topic of ongoing interest in natural language processing (NLP) with
applications in various areas such as authorship classification and spam detection.
Different approaches have been developed, ranging from statistics to machine learning and sentiment analysis
(see e.g.~\cite{MikolovEtAl13},~\cite{PenningtonEtAl14},~\cite{PetersEtAl18},~\cite{Devlin.2019}).
An overview is given in~\cite{Kowsari.2019}.

In our research, we apply common classification methods to numeric vectors derived from text documents by
means of the term frequency-inverse document frequency (\tfidf{})
method (see e.g.~\cite{Jurafsky.2008}, in detail see~\cite{Luhn.1957} for term frequency and for inverse document
frequency~\cite{SPARCKJONES.1972}).
This technique is widely used in NLP.~%
While further developments have been made in this area (e.g.~\cite{Domeniconi.2015}), we concentrate on standard \tfidf{} in this study.
The core idea of the term frequency approach is to create numerical vectors whose individual components
represent essentially the number of occurrences of different character sequences, or "terms".
Complementary, the inverse document frequency concept is directed at assessing the significance of a
word in a text compared to its occurrence in a set of texts.

Text classification needs to deal with several sources of diversity.
In addition to the challenge of imbalance, in our case the number of documents per department,
difficulties may be posed by the existence of different semantic topics per department and variance in text length.
Especially our approach using term frequency may be overly sensitive to differences in text length.
This is due to the fact that the text length will affect the norm of the vectors produced by \tfidf{}.

If e.g.~a short text were simply concatenated with itself several times, all term frequencies would be amplified.
While the text's meaning would then remain unaltered, a classifier trained to take certain term frequencies as the
base for class predictions might arrive at very different results.

The reason is that similar inputs differing only in norm\footnote{The norm of a numeric vector is a value derived from its components which can be interpreted as its magnitude or the distance between the vector's start and end points.}, i.e.~where one vector is a multiple of the other,
will in general not be treated equally by standard classifiers, potentially leading to differences in classification.

One approach to addressing this challenge is to segment or split the data into pieces of equal length, allowing the
classification algorithms to be trained on comparable data.
An alternative strategy involves normalizing the numeric vectors derived from document sets, so that vectors become equal w.r.t.~a chosen norm.
We refer to these two approaches also as the segment and the document based analysis.

In this analysis we deal with those two different ways of mitigating the disturbing effects of varying text lengths.
Our application example is based on the implementation of automatic routing of digital mail in German municipalities.
For these municipalities' administrations a standard layout exists with a maximum of 45 departments, 31 of which are present in our data set.

\section{Research question}\label{sec:research_question}
While data imbalance for classification tasks and different topics per class are well studied, to the  knowledge of the authors, differences in text length is a topic for which no general approach is known. 
Besides this, classification tasks for German text documents are rare.
Therefore, we focus in this study on different approaches to cope with variance in text length when classifying German
documents from public administration.

A comparative study investigating the segment based analysis and applies different aggregation methods to all the results
from different combinations of preprocessing pipelines and classifiers is given in ~\cite{Diez.2023}.
While for the splitting approach an aggregation method for the predicted classes per segment is needed to be able to
classify the whole document, this is not necessary for the normalization approach.

Since the preprocessing of text, including the generation of features, is essential for the model's performance as
highlighted in~\cite{Skiena.2017}, this publication compares the results from prior research presented in~\cite{Diez.2023}
with the approach of normalization.
To ensure consistency and comparability of the findings from both analyses, meticulous
attention was given to utilizing identical experiments throughout our study.

We consider in this paper the \emph{research question}, whether document normalization or segmentation is preferable
for the imbalanced data from German administration.

This work is structured as follows: In Sect.~\ref{sec:datapreproc} we describe the data preprocessing steps at the text level in detail,
followed by a description of the measures taken towards balancing the data.
Sect.~\ref{sec:experiment_design} lays out the experimental setup used to test the different approaches, including a
description of the steps in the classification pipeline.
Finally, we provide our key results in Sect.~\ref{sec:results} and discuss them in Sect.~\ref{sec:consequences}.
\section{Data preprocessing}
\label{sec:datapreproc}
The data for our analysis consists of a corpus of 1,462 text examples. Our project partner Governikus Gmbh \& Co
KG obtained those by collecting pdf documents from municipal web pages, extracting text data and assigning class labels.
Before the segmentation step, some transformations are applied to the data which aim at increasing digestibility for
our classification pipelines, e.g.~removal of punctuation or presumably unspecific terms as well as stemming.
As a result our experiment is based on lists of character sequences, or \emph{terms}, derived from the original texts.

To ensure comparability of the document and segment based analyses, we always perform the segmentation step
and the elimination of segments which is described in~\ref{subsec:databalancing}.
For the normalizing approach, the remaining segments of each document are then concatenated, yielding the texts that
our pipelines were effectively tested on (see Fig.~\ref{img:experiment_pipelines}).

\def\scaler{1.0}
\def\boxheight{\scaler*0.9cm}
\def\boxgap{\scaler*0.4cm}
\def\stdboxwidth{\scaler*2cm}
\tikzstyle{rect} = [rectangle, draw, text centered, text width=\stdboxwidth, minimum height=\boxheight]
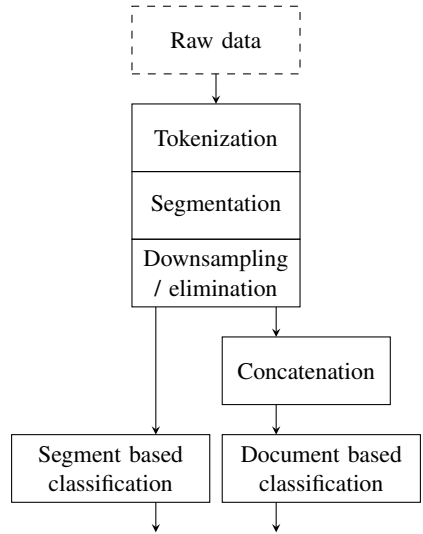
\begin{figure}
		\centering
		\begin{tikzpicture}
			\node[rect, font=\small, dashed]  at (0*\scaler, 0) {Raw data};
			\draw[-stealth] (0*\scaler, -0.5*\boxheight) -- (0*\scaler, -0.5*\boxheight -\boxgap);
			\node[rect, font=\small]  at (0*\scaler, -\boxgap -\boxheight) {Tokenization};
			\node[rect, font=\small]  at (0*\scaler, -\boxgap -2*\boxheight) {Segmentation};
			\node[rect, font=\small]  at (0*\scaler, -\boxgap -3*\boxheight) {Downsampling / elimination};

			\draw[-stealth] (-0.4*\stdboxwidth, -\boxgap -3.5*\boxheight) -- (-0.4*\stdboxwidth, -3*\boxgap -4.5*\boxheight);
			\node[rect, font=\small, text width=1.2*\stdboxwidth]  at (-0.6*\stdboxwidth -0.5*\boxgap, -3*\boxgap -5*\boxheight) {Segment based classification};
			\draw[-stealth] (-0.4*\stdboxwidth, -3*\boxgap -5.5*\boxheight) -- (-0.4*\stdboxwidth, -4*\boxgap -5.5*\boxheight);

			\draw[-stealth] (0.4*\stdboxwidth, -\boxgap -3.5*\boxheight) -- (0.4*\stdboxwidth, -2*\boxgap -3.5*\boxheight);
			\node[rect, font=\small]  at (0.5*\stdboxwidth +0.5*\boxgap, -2*\boxgap -4*\boxheight) {Concatenation};
			\draw[-stealth] (0.4*\stdboxwidth, -2*\boxgap -4.5*\boxheight) -- (0.4*\stdboxwidth, -3*\boxgap -4.5*\boxheight);
			\node[rect, font=\small, text width=1.2*\stdboxwidth]  at (0.6*\stdboxwidth +0.5*\boxgap, -3*\boxgap -5*\boxheight) {Document based classification};
			\draw[-stealth] (0.4*\stdboxwidth, -3*\boxgap -5.5*\boxheight) -- (0.4*\stdboxwidth, -4*\boxgap -5.5*\boxheight);
		\end{tikzpicture}

	\caption{Preprocessing steps at the text level which produce the input for the segment and document based classification pipelines.}
	\label{img:experiment_pipelines}
\end{figure}
Subsection~\ref{subsec:textual_preproc} describes how texts are converted to lists of terms and then to segments.
Imbalance in the class sizes is the topic of subsection~\ref{subsec:databalancing}.

\subsection{Textual preprocessing }\label{subsec:textual_preproc}
The raw data extracted from a pdf documents contains a lot of information which may be unhelpful to a bag of word based
	analysis.
This includes terms which are unlikely to highly relate to any specific class, such as first names or place names,
but also typical endings introduced by adapting words to their function in a sentence, as is typical for a moderately
highly inflected language such as German.
This section deals with all the changes performed at the text level, before creating the first numeric vectors.\\

First, every text is reduced to German letters and digits by replacing every sub-string with a blank which consists exclusively of characters that do not belong in any of these categories.
In what follows, we consider every character sequence that contains no blanks a "token".
Tokens which represent known inflected forms of words appearing in a certain lemmatization dictionary with  447,755 known shapes are then replaced with their canonical form, or alternatively a different root form as indicated by that dictionary.
This is supposed to facilitate recognition of words independently of context-related inflection effects.
Next, all tokens are removed which have less than three different characters or contain only digits, as well as all those appearing in given lists of stop words, place names and first names. 
The CISTEM (\cite{Weissweiler}) stemmer for the German language, which strips off some common prefixes and suffixes,
is used to further reduce related words to common root forms.
Finally, the texts are converted to lowercase and all tokens that do not consist of letters only are removed.\\

At this point, every text has been transformed into a string containing a blank-separated array of sequences of German letters.
These strings are now turned into segments of 2\,048 characters each.
Usually, there will be a shorter segment at the end.

Some of these segments are removed as part of the undersampling effort described in~\ref{subsec:databalancing}.
For the document based analysis, the remaining segments of a document are concatenated.

By cutting unconditionally after a fixed number of characters, without regard for word boundaries, we sometimes split
tokens.
Thus, some new character sequences are created accidentally.
The character sequences which are present now are the terms that are considered in the creation of term frequency
vectors which marks the transition to numeric vectors (see~\ref{subsec:preprocess_to_classification}).

When concatenating the segments for the document based analysis, blanks are inserted at the joints to insure that the
remaining terms for each document are the same in both the segment and the document based classification.

\subsection{Data balancing}\label{subsec:databalancing}
The splitting procedure described in~\ref{subsec:textual_preproc} delivered a total of 15,802 text samples, which are distributed across the individual departments very unevenly.
To avoid the potentially damaging effect of imbalanced data, \cite{Diez.2023} used elimination of segments and reduction of the classes considered. Note, this eliminates no documents completely. We also follow the restriction on classes with at least 100 segments, which leaves 31 classes in the pool.
As mentioned, for comparability, we follow the same steps here.
\begin{figure}[ht]
	\includegraphics[width=\textwidth, page=2]{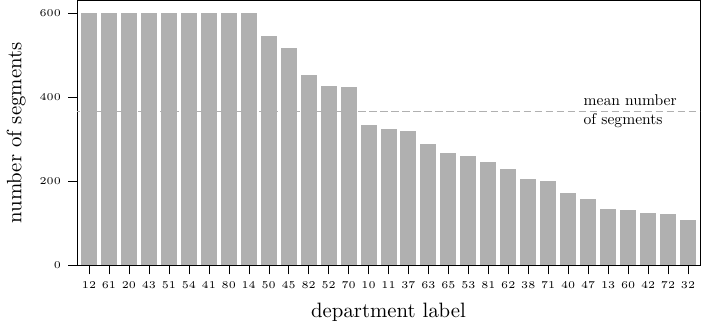}
	\caption{Number of segments by department after filtering.}
	\label{fig:hist_filtered_segs}
\end{figure}
\begin{figure}[ht]
	\includegraphics[width=\textwidth, page=3]{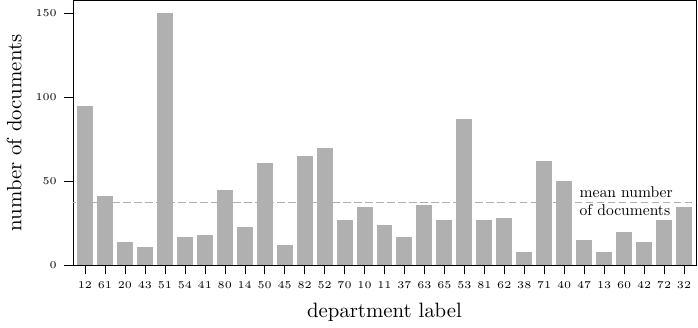}
	\caption{Number of documents by department after filtering.}
	\label{fig:hist_filtered_docs}
\end{figure}
Finally, 11,386 segments stemming from 1,169 documents remain.
The number of segments per department starts from 107 and averages 367.3,
while for documents the minimum is 8 documents and the average 37.7,
as illustrated by Figs.~\ref{fig:hist_filtered_segs} and~\ref{fig:hist_filtered_docs}.\footnote{Corrected sample standard deviations
are 187.5 for the segment and 31.0 for the document view.}
All pipelines used in the analysis contain an oversampling step (see~\ref{subsec:preprocess_to_classification}).
In the segment based analysis it increases the number of samples in all smaller classes to the size of the
majority class.
Were this step to be performed on the data set as described, the result would contain 38.8 \% generated data.
The oversampling step is, however, part of the classification pipeline which is applied in a 5-fold cross validation
(CV) setting.
As the classes exhibit larger variance in the folds, we end up with 47.9 \% generated data.

As mentioned, there is relatively more variance in class sizes in the document based analysis, so that oversampling to
the majority class' size would have led to a considerably larger share of generated data.
We therefore restrict the oversampling to 55 observations per class and leave larger classes unaltered.
That way, the share of generated data is, similarly, 38.8 \% in the total data set and 48,3 \% when considering CV
folds, at the cost of retaining some unevenness in class sizes.

\section{Experiment Design - Pipeline}\label{sec:experiment_design}
In this section, we will discuss the design of our experiment, which aims to compare the performance of various classification algorithms on pre-processed texts. The study will focus on evaluating the effectiveness of different algorithms in accurately categorizing texts into predefined categories.
We describe the different algorithms we use, the constructed versions of the pipeline and which evaluation technique we use.

\subsection{Preprocess for classification}
\label{subsec:preprocess_to_classification}

For both the segment based and the document based classification, we apply the same preprocessing steps as in \cite{Diez.2023} (see Picture~\ref{fig:experiment_pipelines}).
These are
\begin{enumerate}
	\item Term count vectorization,
	\item L1-normalization,
	\item Oversampling and
	\item \tfidf{} converter.
\end{enumerate}

First, we transform each term into numeric.
This is the basis, so that classification algorithms can be trained.
As stated, we consider \emph{bag of words} classification models, where not the relation between terms is of importance, but simply their frequency,
i.e.~we use term frequency (TF, for details see~\cite{Luhn.1957}).

In order to uniquely represent the terms present in the document, it is essential to construct a dictionary.
Each term in the dictionary is analyzed for its frequency of occurrence within the document.
This analysis involves representing each document as a vector of equal length, corresponding to the unique terms identified in the dictionary.
By encoding each term as a binary vector, where only one position is designated as one and all others are zero, a more streamlined representation is achieved.
Combining these binary vectors for each term within a document results in a term frequency (TF) vector.
The TF vector records the count number of each term present in the document, with zero indicating terms that do not appear.
Note that semantic information is not preserved.
The L1 norm of the vector reflects the total number of terms in the document.
Stacking these TF vectors from all documents generates a term frequency matrix.
In this matrix, each row corresponds to a specific document, while each column represents an individual term.
This matrix provides a comprehensive overview of term frequencies across all documents analyzed.

All pipelines in the analysis contain an L1 normalizer followed by an oversampling step (see~\ref{subsec:pipeline}).
In order to understand the application of the L1 normalizer, it is important to first delve into the functionality of the Oversampler. The numerical oversampling process mentioned in Section~\ref{subsec:databalancing} is executed using the SMOTE method introduced by Chawla et al. (2002)~\cite{Chawla.2002}.
This method involves generating new items by randomly selecting one element from the same class, followed by choosing one of its five nearest
neighbors.\footnote{For document based analysis, the number of neighbors considered was reduced to four, as not all classes contain more than five members.}
The new vector is created as a convex combination, with the coefficients determined by another random element.
To ensure that the sampled element accurately represents the class, all document vectors are scaled to fit within the same hypersphere.

Observe, the oversampler needs to be treated differently for segmentation based and normalized document based pipelines, as mentioned in \ref{subsec:databalancing}.\\

Up to this point, all terms have been treated as equally important.
As very common terms can be expected to contribute less helpful information than rather specific ones, it makes sense
to multiply the term frequencies by a value which increases with the specificity of terms. Such a value is given by the so-called inverse of the document frequency (IDF~\cite{SPARCKJONES.1972} ), i.e. the inverse of the number of documents a term appears in or some derivate thereof.
In our pipelines we weight the columns in the TF matrix with the logarithm of the inverse fraction of documents the respective term appears in.

\subsection{Pipeline}\label{subsec:pipeline}

Since we compare our results for normalized documents with the results form our previous work (~\cite{Diez.2023}), we use the same pipelines.

Prior to applying classification algorithms, it is essential to conduct various preprocessing steps based on the provided texts. While numerous preprocessing techniques are conceivable, we focus on four specific types of preprocessing steps tailored to the insights gained in Section ~\ref{subsec:preprocess_to_classification}.
These include
\begin{itemize}
	\item Converter: Transforming text into a vector of token counts or term counts into term frequency-inverse document frequency (\tfidf{}).
	\item Normalizer: Scaling vectors to equal L1 or L2 norm.
	\item Oversampler: Implementing synthetic minority oversampling.
	\item Input dimension reduction: Utilizing truncated singular value decomposition.
\end{itemize}

Following the framework proposed in~\cite{Diez.2023}, we explore four distinct experiment pipelines incorporating these preprocessing steps, as illustrated in Figure~\ref{img:experiment_pipelines}. These pipelines differ from segmentation approaches by omitting the splitting and aggregation step before and after the pipeline, respectively. Instead, an L1 normalization is applied post-term frequency transformation to accommodate variations in document lengths.
For details on the sequencing of steps, particularly the L1 normalization, we refer to Section~\ref{subsec:preprocess_to_classification}. 
\def\scaler{1.0}
\tikzstyle{rect} = [rectangle, draw, text centered, text width=6em, minimum height=3em]
\begin{figure}
	\begin{minipage}{0.22\linewidth}
		\centering
		\emph{First Pipeline}
		\hspace{3em}
		
		\begin{tikzpicture}
		\draw[-stealth] (0*\scaler,0*\scaler) -- (0*\scaler,-2em*\scaler);
		\node[rect, font=\small]  at (0*\scaler, -3.5em*\scaler) {Token count converter};
		\node[rect, font=\small]  at (0*\scaler, -6.5em*\scaler) {L1 normalizer};
		\node[rect, font=\small]  at (0*\scaler, -9.5em*\scaler) {Oversampler};
		\node[rect, font=\small]  at (0*\scaler, -12.5em*\scaler) {\tfidf{} converter};
		\node[rect, font=\small]  at (0*\scaler, -15.98em*\scaler) {Truncated singular value decomposition};
		\node[rect, font=\small]  at (0*\scaler, -19.47em*\scaler) {L2 normalizer};
		\node[rect, font=\small]  at (0*\scaler, -22.48em*\scaler) {Classifier};
		\draw[-stealth] (0*\scaler,-23.98em*\scaler) -- (0*\scaler,-25.98em*\scaler);
		\end{tikzpicture}
		
	\end{minipage}
	\begin{minipage}{0.22\linewidth}
		\centering
		\emph{Second Pipeline}
		\hspace{3em}
		
		\begin{tikzpicture}
		\draw[-stealth] (0*\scaler,0*\scaler) -- (0*\scaler,-2em*\scaler);
		\node[rect, font=\small]  at (0*\scaler, -3.5em*\scaler) {Token count converter};
		\node[rect, font=\small]  at (0*\scaler, -6.5em*\scaler) {L1 normalizer};
		\node[rect, font=\small]  at (0*\scaler, -9.5em*\scaler) {Oversampler};
		\node[rect, font=\small]  at (0*\scaler, -12.5em*\scaler) {\tfidf{} converter};
		\node[rect, font=\small]  at (0*\scaler, -15.98em*\scaler) {Truncated singular value decomposition};
		
		\draw[-stealth] (0*\scaler,-17.97em*\scaler) -- (0*\scaler,-20.97em*\scaler);
		\node[rect, font=\small]  at (0*\scaler, -22.48em*\scaler) {Classifier};
		\draw[-stealth] (0*\scaler,-23.98em*\scaler) -- (0*\scaler,-25.98em*\scaler);
		\end{tikzpicture}
	\end{minipage}
	\begin{minipage}{0.22\linewidth}
		\centering
		\emph{Third Pipeline}
		\hspace{3em}
		
		\begin{tikzpicture}
		\draw[-stealth] (0*\scaler,0*\scaler) -- (0*\scaler,-2em*\scaler);
		\node[rect, font=\small]  at (0*\scaler, -3.5em*\scaler) {Token count converter};
		\node[rect, font=\small]  at (0*\scaler, -6.5em*\scaler) {L1 normalizer};
		\node[rect, font=\small]  at (0*\scaler, -9.5em*\scaler) {Oversampler};
		\node[rect, font=\small]  at (0*\scaler, -12.5em*\scaler) {\tfidf{} converter};
		
		\draw[-stealth] (0*\scaler,-14.04em*\scaler) -- (0*\scaler,-18em*\scaler);
		\node[rect, font=\small]  at (0*\scaler, -19.47em*\scaler) {L2 normalizer};
		\node[rect, font=\small]  at (0*\scaler, -22.48em*\scaler) {Classifier};
		\draw[-stealth] (0*\scaler,-23.98em*\scaler) -- (0*\scaler,-25.98em*\scaler);
		\end{tikzpicture}
	\end{minipage}
	\begin{minipage}{0.22\linewidth}
		\centering
		\emph{Fourth Pipeline}
		\hspace{3em}
		
		\begin{tikzpicture}
		\draw[-stealth] (0*\scaler,0*\scaler) -- (0*\scaler,-2em*\scaler);
		\node[rect, font=\small]  at (0*\scaler, -3.5em*\scaler) {Token count converter};
		\node[rect, font=\small]  at (0*\scaler, -6.5em*\scaler) {L1 normalizer};
		\node[rect, font=\small]  at (0*\scaler, -9.5em*\scaler) {Oversampler};
		\node[rect, font=\small]  at (0*\scaler, -12.5em*\scaler) {\tfidf{} converter};
		
		\draw[-stealth] (0*\scaler,-14.04em*\scaler) -- (0*\scaler,-21em*\scaler);
		\node[rect, font=\small]  at (0*\scaler, -22.48em*\scaler) {Classifier};
		\draw[-stealth] (0*\scaler,-23.98em*\scaler) -- (0*\scaler,-25.98em*\scaler);
		\end{tikzpicture}
	\end{minipage}
	\caption{The four pipelines dealing with different document lengths by L1 normalization.}
	\label{fig:experiment_pipelines}
\end{figure}
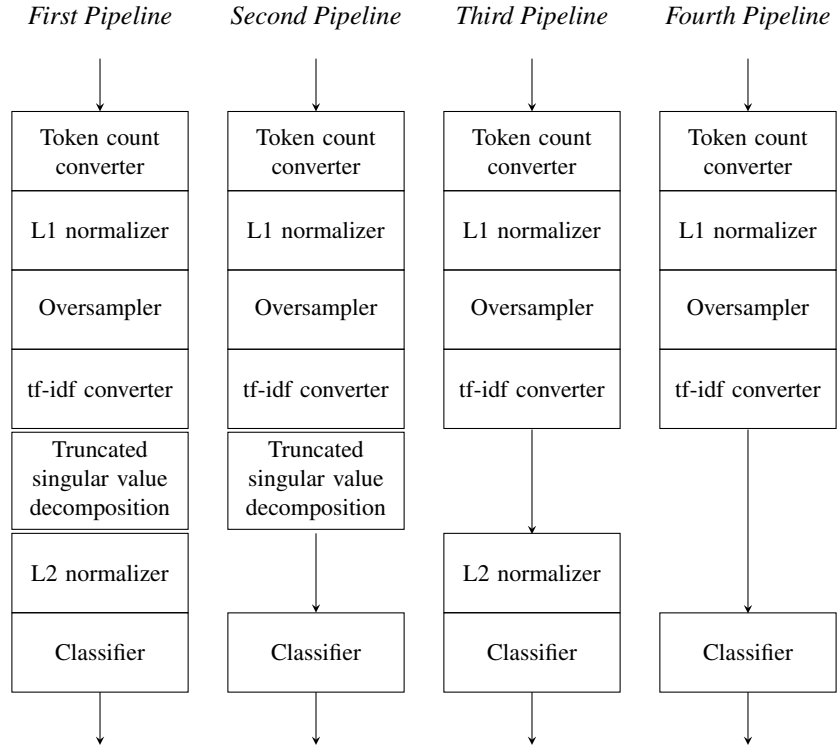

 The differences between the four pipelines are in the dimensionality reduction and the L2 normalizer.
 In the first and second pipelines, dimension reduction is implemented using truncated singular value decomposition to reduce the dimensionality of
 the \tfidf{} matrix to 800.
 This reduction forms the basis for L2 normalization in the first pipeline.
 The second pipeline omits scaling, while the third pipeline scales \tfidf{} vectors without performing dimension reduction.
In contrast, the fourth pipeline directly classifies documents based on their \tfidf{} vectors without additional preprocessing steps.
The outcome of each pipeline is then the input for the classifier.

\subsection{Classification algorithm}\label{sec:classification_algorithm}
In this section, we focus on exploring various classifiers commonly used in machine learning. The classifiers considered in this study include
\begin{itemize}
	\item logistic regression (\logisticregression, see e.g.~\cite{Hastie.2009}),
	\item random forest (\randomforest, see e.g.~\cite{Breiman.1984}),
	\item support-vector machine (\supportvectorclassifier, see e.g.~\cite{Cortes.1995}),
	\item feed-forward neural networks (\neuralnet, see e.g.~\cite{Goodfellow.2016}), and
	\item supervised variational autoencoder (\SVAE, see e.g.~\cite{Quint.2018}).
\end{itemize}

Each of these classifiers comes with hyperparameters that need to be set prior to training.
To optimize the performance of the models, we conduct a hyperparameter optimization process.
We employ a hyperparameter optimization technique known as Bayesian hyperparameter search (refer to~\cite{Mockus.1989})
which creates different combinations of hyperparameters for each algorithm, optimizes the model for each set and select
the configuration that yields the highest accuracy.

The specific hyperparameter search space for each classifier is detailed in Table~\ref{tab:hyperparameter_search_space}.

\begin{table}
	\caption{Hyperparameter search space.}
	\label{tab:hyperparameter_search_space}
	\begin{tabular*}{\linewidth}{llp{4.2cm}}
		\hline\noalign{\smallskip}
		Classifier & Hyperparameter & Search Space \\	\noalign{\smallskip}\hline\noalign{\smallskip}
		\logisticregression & Regularization parameter & [1e-6; 100] \\
		& Penalty type & \{l1; l2; elasticnet; no\}\\
		& Elastic-Net mixing parameter & [0; 1]\\
		& Tolerance for stopping criteria & [1e-6; 1e-2] \\[3mm]
		\neuralnet & Number of hidden layers & \{1; 2; 3\} \\
		& Hidden layer size & [1; 500] \\
		& Activation function & \{logistic; tanh; relu\} \\
		& Initial learning rate & [1e-6; 1e-2] \\
		& Tolerance of optimizer & [1e-6; 1e-2] \\
		& Maximum number of epochs to not meet tolerance improvement & [1; 100] \\[3mm]
		\randomforest & Number of trees & [1; 1,000] \\
		& Maximum tree depth & [1; 1,000] \\[3mm]
		\SVAE & Layer number encoder & \{1; 2; 3\}\\
		& First layer size encoder & [10; 500]\\
		& Follow-up encoder layer sizes relative to previous & [0.001; 0.9]\\
		& Latent dimension relative to 1st encoder layer & [0.001; 0.9]\\
		& Loss weight variational autoencoder & [1.0; 10.0]\\
		& Loss weight classifier & [1.0; 10.0]\\
		& Activation function & \{logistic; relu; tanh;\newline\hspace*{1em} sigmoid\}\\
		& Tolerance of optimizer & [1e-6; 1e-2] \\
		& Maximum number of epochs to not meet tolerance improvement & [1; 100]\\
		& Maximum number of epochs & [1; 100] \\[3mm]
		\supportvectorclassifier & Regularization parameter & [1e-6; 100] \\
		& Kernel & \{rbf; linear\}\\
		& Kernel coefficient & [1e-6; 1e-2] \\
		& Tolerance for stopping criteria & [1e-6; 1e-2] \\
		\noalign{\smallskip}\hline\noalign{\smallskip}
	\end{tabular*}
\end{table}

\subsection{Evaluation technique}\label{sec:evaluation}
In order to compare the performance of the models, key statistical metrics including \emph{Accuracy} (Acc.), \emph{Precision} (Prec.), \emph{Recall} (Rec.), and \emph{F1-Score} (F1) are computed using data from all test sets. These metrics provide valuable insights into the effectiveness of the classification models.
\begin{itemize}
	\item \emph{Acc.} represents the ratio of correctly classified documents across all classes.
	\item \emph{Prec.} measures the proportion of observations assigned to a class that truly belong to that class.
	\item \emph{Rec.} is determined by dividing the number of correctly identified members of a class by the total number of true class members. In a multiclass setting where sample sizes per class vary, the class-weighted average recall aligns with the accuracy statistic.
	\item The \emph{F1} for each class is calculated as the harmonic mean of precision and recall.
\end{itemize}
To ensure a comprehensive evaluation in the multiclass framework with varying sample sizes per class, Prec., Rec.,
and F1 are derived as averages of per-class evaluations weighted based on class size.
This approach accounts for the imbalanced distribution of samples across different classes and provides a more robust assessment of model performance.

In our earlier work~(\cite{Diez.2023}) investigating the segment based analysis, we tested an array of aggregation methods
to predict the class of a document based on the predictions probabilities for all of its segments.
The evaluation for the segment based analysis involves calculating the same statistical metrics, based on the predictions
produced by the aggregation methods.
In that analysis we concluded that the same three aggregation algorithms always occupied the leading positions, namely
\begin{itemize}
	\item \maxsumprob{} (\maxsumprobAc),
	\item \maxweightedavgprob{} (\maxweightedavgprobAc) and
	\item \summaxprob{} (\summaxprobAc).
\end{itemize}
In our comparative study, we used the three aggregation variants for the segment based results, as described in Section~\ref{sec:results}.

\subsection{Conducting the experiment}
For our experiments, we implemented pipelines including the preprocessing in the Python programming language.
Many components from scikit-learn~(\cite{sklearn}) and imbalanced-learn~(\cite{imblearn}) went into this implementation.

We consider the four different pipelines outlined in \ref{subsec:pipeline}. 
All of these are paired with each of the classifiers mentioned in ~\ref{sec:classification_algorithm}.

We adopt a five-fold cross-validation strategy to evaluate the models.
To maintain consistency and enable comparison with previous results based on segment analysis (as presented in ~\cite{Diez.2023}),
we use the same folds as in that study.
Those folds were constructed with the restriction to keep all segments of a document together within each fold even for the
segment based analysis, while distributing the segment numbers across the folds as evenly as possible.\footnote{The maximum difference is one.} 
Each partition is used once as a test set, while the remaining subsets collectively serve as the training set for model training and evaluation.

Since advantageous hyperparameters can significantly enhances model performance, we consider their choice in setting up our experiment.
Prior to the evaluation, we thus perform a hyperparameter search as described in~\ref{sec:classification_algorithm}.
The same cross-validation folds used for evaluation purposes (detailed in Section ~\ref{sec:evaluation}) are also employed
in this optimization process, and the configurations are evaluated across all test folds.
This approach ensures the robustness and generalizability of the models.

For each model, endowed with the best found set of hyperparameters, the prediction results from all five test folds form
the basis for the evaluation.
In the segment based case, document predictions are derived using the most successful aggregation methods from~\cite{Diez.2023}.
Finally, evaluation scores are calculated according to~\ref{sec:evaluation}.

We conducted the experiments on CPUs
 except for the variational autoencoder which used Nvidia A100 GPUs.

\section{Results}\label{sec:results}
We present the results in tables showing the best hyperparameter values determined by the  Bayes optimizer, table \ref{tab:hyper_parameter},
and the corresponding results for the pipelines, tables \ref{tab:results_of_experiment_1_2} and \ref{tab:results_of_experiment_3_4}.

We first shortly analyze the algorithms w.r.t. the best hyperparameters and conclude afterwards the results w.r.t. the preprocessing choice of segmentation or normalization.

Only for SVM we see a clear preference of linear kernel function for all four pipelines. In other algorithms, a preference of some of the hyperparameters can be observed, e.g. compare the activation function for SVAE, the penalty for LR, the number of hidden layer for NN and RF tends to max out the number of trees.

For most experiments, the document based classifiers outperform their segment based counterparts in terms of accuracy values, with the exception
of pipelines 2, 3, 4 with the SVAE classifier and pipelines 3 and 4 with the SVM classifier.

For pipeline 1, the advantage of the document based runs ranges from 0.68 for the SVAE up to 4.19 percentage points for the RF,
as shown in table~\ref{tab:results_of_experiment_1_2}.
The RF is the worst classifier out of all used classifiers.
The best aggregation method for the segmentation is the MWA, second only to RMS for the NN and tied with MS for the SVM.

Results for pipeline two are in table~\ref{tab:results_of_experiment_1_2}.
The advantage of the document based experiments varies more widely and peaks at 9.16 in percentage points for the RF, which is again the worst classifier.
For SVAE, the segment based experiment performs even slightly better, at 1.11 percentage points.m
There exists no unequivocally best aggregation method for this pipeline.

With pipeline three, shown in table~\ref{tab:results_of_experiment_3_4}, the segment experiment are better for the SVAE and SVM,
with less than 6.42 and 4.11 percentage points, respectively.
On the other hand, the highest lead of the document side is 1.62 percentage points with the LR.
The best aggregation method for the pipeline is again MWA, except for the NN.
This pipeline features the overall best accuracy value from segment based experiments, at 88.45 \%  when using  RMS.

Pipeline 4 shows the smallest differences between segment and documentwise classification among all pipelines.
For this pipeline, no aggregation method has an advantage.

Overall, we can conclude that for pipelines containing the dimensionality reduction, mainly all algorithms perform worse in the segmentation scheme, than in the pipelines without reduction, except NN and RF. 
\section{Discussion and Summary}\label{sec:consequences}
In conclusion, our study highlights the importance of normalization techniques in text analysis tasks.
We compared two approaches of handling different text lengths, the segmentation and normalization methods.
Based on our results we can conclude the following:
\begin{itemize}
	\item Segmentation should be considered carefully for text analysis.
	\item For SVM and SVAE classifiers the L2-normalization should not be left out.
	\item The logistic  regression and neural network should always be considered in text classification.
\end{itemize} 
The first statement is based on the observation that the implementation of normalization techniques and utilization of
complete information results in enhanced classification outcomes.
Nevertheless, the segmentation process introduces a challenge as the information contained within segments may lack comprehensive
details, resulting in heightened variability within segments and an elevated risk of misclassification.
Segmentation inherently covers less information as the vector of a segment contains a smaller number of positive values
compared to the vector of a document.
The dimension reduction further reduces this information, potentially impacting classification outcomes.
It is essential to consider these limitations and potential effects on classification accuracy when utilizing segmentation
in text analysis tasks.
These results underscore the importance of carefully considering segmentation strategies to ensure optimal classification
accuracy in text analysis tasks.

We found that inclusion of the L2 normalization step consistently delivers superior results for SVM and SVAE
regardless of whether it is applied to documents or segmentation.
In SVM, the use of the Euclidean norm without scaling inadvertently assigns higher weights to certain features over others, potentially skewing classification outcomes.
For SVAE, maintaining consistent mean and variance values across batches is crucial to ensure optimal performance.
This result emphasizes the critical role of normalization techniques in enhancing the effectiveness and reliability of SVM and SVAE models in text analysis tasks.

Based on the results, the choice of classifier is not significant in terms of the performance criteria considered compared to the selection of pipelines.
We recommend considering logistic regression in every analysis as it consistently delivers good performance and has a simple structure.
Additionally, we suggest using neural networks for comparison purposes as they exhibited the best performance in most pipelines.

In this study we focused on classification models while we compared different approaches of preprocessing to handle the different text lengths.
We set a fixed length for the segmentation approach,
yet the segment length might have a significant effect on the performance outcomes.
The results we observed and conclusion we drew may consequently have been influenced by that choice.
Therefore, it remains an unanswered question as to which extent varying the segment length could yield improved performance results.

An alternative approach could be a combination of methods.
Given the diversity of the document lengths present in our dataset, a combination of the methods to treat the documents differently
depending on their length might lead to even better results.

In addition to the classification algorithms tested here, there are more advanced text processing methods
that could be explored in future research.
Moving forward, our next objective is to compare some of these methods with the approaches utilized in this study.

\section*{acknowledgement}
	This work was supported by the Ministry of Education and Research Germany (BMBF, grant number 01IS20061 ("EP-KI")).

	The company Governikus Gmbh \& Co KG provided the data and the textual preprocessing steps as well as valuable insight into
	the administrative challenge of content based routing.

\clearpage
\section*{Appendix 1: Results in Detail}\label{sec:results_in_detail}
\addcontentsline{toc}{section}{Appendix 1: Results in Detail}
\begin{table}
	\caption{Best-performing hyperparameter sets for all experiments.}
	\label{tab:hyper_parameter}
	\begin{tabular*}{\linewidth}{llllllllll}
		\hline\noalign{\smallskip}
		Classifier & Hyperparameter & \multicolumn{4}{l}{Segment} & \multicolumn{4}{l}{Document} \\
		&  & 1 & 2 & 3 & 4 & 1 & 2 & 3 & 4 \\
		\noalign{\smallskip}\hline\noalign{\smallskip}
		\SVAE & \clfactivation & tanh & tanh & tanh & sigmoid & tanh & tanh & sigmoid & relu \\
		& \clfencoderdim & 500 & 286 & 469 & 500 & 241 & 155 & 468 & 500 \\
		& \clflatentdim & 8.33e-2 & 73.6e-2 & 5.86e-2 & 27.9e-2 & 25.0e-2 & 44.5e-2 & 21.2e-2 & 6.75e-2 \\
		& \clfniternochange & 1 & 1 & 63 & 1 & 70 & 55 & 84 & 36 \\
		& \clfratioi & none & 78.1e-2 & none & none & 59.6e-2 & none & 35.5e-2 & 0.9 \\
		& \clfsampleweightclf & 10 & 10 & 10 & 10 & 4.45 & 9.70 & 7.90 & 9.30 \\
		& \clfsampleweightvae & 1 & 2.78 & 1 & 1 & 5.24 & 1 & 1 & 1 \\
		& \clftol & 1e-2 & 8.8e-5 & 1.15e-4 & 1e-2 & 1.5e-5 & 6.61e-4 & 2.79e-3 & 2e-6 \\[1mm]
		\logisticregression & \clfC & 14.2 & 8.75e-3 & 9.16e-4 & 6.89 & 8.86e-3 & 66.60 & 100 & 100 \\
		& \clfpenalty & none & l1 & none & none & none & l1 & none & l2 \\
		& \clftol & 9.07e-3 & 5.7e-4 & 9.5e-4 & 2.9e-4 & 1.9e-5 & 1e-6 & 1e-6 & 1.40e-4 \\[1mm]
		\neuralnet & \clfactivation & logistic & logistic & tanh & logistic & tanh & relu & tanh & tanh \\
		& \clflayeri & 383 & 286 & 236 & 336 & 255 & 487 & 320 & 307 \\
		& \clflayerii & none & none & 139 & 470 & none & none & none & none \\
		& \clflayeriii & none & none & 347 & none & none & none & none & none \\
		& \clflearningrateinit & 6.6e-5 & 1e-6 & 4.44e-4 & 4.4e-5 & 2.06e-4 & 5.52e-3 & 1e-2 & 1e-2 \\
		& \clfniternochange & 43 & 100 & 39 & 11 & 76 & 74 & 100 & 100 \\
		& \clftol & 9.07e-4 & 1.2e-5 & 7.6e-5 & 1e-2 & 2.74e-3 & 1.81e-3 & 1e-6 & 7.88e-3 \\[1mm]
		\randomforest & \clfmaxdepth & 308 & 984 & 814 & 706 & 61 & 258 & 1000 & 1000 \\
		& \clfnestimators & 916 & 932 & 830 & 878 & 878 & 974 & 1000 & 602 \\[1mm]
		\supportvectorclassifier & \clfC & 0.695 & 1.67e-3 & 3.89 & 4.15e-4 & 1.58 & 32.4 & 63.5 & 100 \\
		& \clfgamma & 1e-2 & 1e-2 & 6.38e-3 & 1e-6 & 1e-2 & 1e-6 & 1e-6 & 1e-6 \\
		& \clfkernel & linear & linear & linear & linear & linear & linear & linear & linear \\
		& \clftol & 1e-2 & 2.08e-3 & 5.94e-3 & 1e-2 & 1e-2 & 1e-6 & 1e-2 & 1e-2 \\
		\noalign{\smallskip}\hline\noalign{\smallskip}
	\end{tabular*}
\end{table}

\begin{table}
	\setlength{\tabcolsep}{0.5em}
	\caption{Results in percentages for pipelines 1 and 2.}
	\label{tab:results_of_experiment_1_2}
	\begin{tabular}{lllllllllll}
		\hline\noalign{\smallskip}
		Base & Classif. & \aggregatemethod& \multicolumn{4}{l}{Pipeline 1} & \multicolumn{4}{l}{Pipeline 2} \\
		&          &                 & Acc. & \precisionweightedAc & \recallweightedAc & \fscoreweightedAc & Acc.
		& \precisionweightedAc & \recallweightedAc & \fscoreweightedAc\\
		\noalign{\smallskip}\hline\noalign{\smallskip}
		\expiai    & \SVAEonelayer            & \maxsumprobAc         & 85.63 & 85.83 & 85.63 & 85.43 & 84.94 & 84.93 & 84.94 & 84.74 \\
		\expiai    & \SVAEonelayer            & \maxweightedavgprobAc & 85.89 & 85.98 & 85.89 & 85.66 & 84.69 & 84.66 & 84.69 & 84.44 \\
		\expiai    & \SVAEonelayer            & \summaxprobAc         & 85.63 & 85.83 & 85.63 & 85.46 & 84.94 & 84.92 & 84.94 & 84.75 \\
		\expiaidoc & \SVAEtwolayer            & none                  & 86.57 & 86.14 & 86.57 & 86.14 & 83.83 & 84.39 & 83.83 & 83.54 \\[1.1mm]
		\expiai    & \logisticregression      & \maxsumprobAc         & 86.66 & 87.07 & 86.66 & 86.58 & 85.12 & 85.56 & 85.12 & 85.04 \\
		\expiai    & \logisticregression      & \maxweightedavgprobAc & 87.17 & 87.52 & 87.17 & 87.05 & 85.29 & 85.67 & 85.29 & 85.17 \\
		\expiai    & \logisticregression      & \summaxprobAc         & 86.40 & 86.83 & 86.40 & 86.32 & 85.54 & 86.04 & 85.54 & 85.47 \\
		\expiaidoc & \logisticregression      & none                  & 89.73 & 89.83 & 89.73 & 89.54 & 88.11 & 88.61 & 88.11 & 87.87 \\[1.1mm]
		\expiai    & \neuralnetonelayerCPU    & \maxsumprobAc         & 87.34 & 87.57 & 87.34 & 87.22 & 87.00 & 87.36 & 87.00 & 86.85 \\
		\expiai    & \neuralnetonelayerCPU    & \maxweightedavgprobAc & 87.51 & 87.67 & 87.51 & 87.32 & 87.25 & 87.61 & 87.25 & 87.11 \\
		\expiai    & \neuralnetonelayerCPU    & \summaxprobAc         & 87.68 & 87.87 & 87.68 & 87.53 & 87.00 & 87.36 & 87.00 & 86.85 \\
		\expiaidoc & \neuralnetonelayerCPU    & none                  & 89.91 & 89.95 & 89.91 & 89.64 & 88.28 & 88.99 & 88.28 & 88.11 \\[1.1mm]
		\expiai    & \randomforest            & \maxsumprobAc         & 74.76 & 78.71 & 74.76 & 73.85 & 72.97 & 78.95 & 72.97 & 72.51 \\
		\expiai    & \randomforest            & \maxweightedavgprobAc & 75.45 & 79.45 & 75.45 & 74.66 & 73.99 & 78.91 & 73.99 & 73.36 \\
		\expiai    & \randomforest            & \summaxprobAc         & 74.25 & 78.31 & 74.25 & 73.61 & 72.54 & 78.16 & 72.54 & 72.11 \\
		\expiaidoc & \randomforest            & none                  & 79.64 & 82.26 & 79.64 & 79.26 & 83.15 & 84.08 & 83.15 & 83.16 \\[1.1mm]
		\expiai    & \supportvectorclassifier & \maxsumprobAc         & 86.48 & 87.06 & 86.48 & 86.42 & 82.38 & 84.99 & 82.38 & 82.57 \\
		\expiai    & \supportvectorclassifier & \maxweightedavgprobAc & 86.48 & 86.98 & 86.48 & 86.36 & 82.98 & 84.97 & 82.98 & 83.06 \\
		\expiai    & \supportvectorclassifier & \summaxprobAc         & 86.23 & 86.91 & 86.23 & 86.17 & 82.55 & 85.23 & 82.55 & 82.90 \\
		\expiaidoc & \supportvectorclassifier & none                  & 89.22 & 89.57 & 89.22 & 88.88 & 85.37 & 86.47 & 85.37 & 85.10 \\
		\noalign{\smallskip}\hline\noalign{\smallskip}
	\end{tabular}
\end{table}

\begin{table}
	\setlength{\tabcolsep}{0.5em}
	\caption{Results in percentages for pipelines 3 and 4.}
	\label{tab:results_of_experiment_3_4}
	\begin{tabular}{lllllllllll}
		\hline\noalign{\smallskip}
		Base & Classif. & \aggregatemethod& \multicolumn{4}{l}{Pipeline 3} & \multicolumn{4}{l}{Pipeline 4} \\
		     &          &                 & Acc. & \precisionweightedAc & \recallweightedAc & \fscoreweightedAc & Acc.
		                                                  & \precisionweightedAc & \recallweightedAc & \fscoreweightedAc\\
		\noalign{\smallskip}\hline\noalign{\smallskip}
		\expib    & \SVAEonelayer            & \maxsumprobAc         & 86.83 & 87.23 & 86.83 & 86.56 & 85.89 & 86.28 & 85.89 & 85.69 \\
		\expib    & \SVAEonelayer            & \maxweightedavgprobAc & 87.17 & 87.57 & 87.17 & 86.90 & 85.54 & 86.04 & 85.54 & 85.38 \\
		\expib    & \SVAEonelayer            & \summaxprobAc         & 86.66 & 87.11 & 86.66 & 86.38 & 85.54 & 85.95 & 85.54 & 85.37 \\
		\expibdoc & \SVAEtwolayer            & none                  & 80.75 & 80.59 & 80.75 & 80.19 & 83.58 & 83.95 & 83.58 & 83.25 \\[1.1mm]
		\expib    & \logisticregression      & \maxsumprobAc         & 87.94 & 88.54 & 87.94 & 87.75 & 86.31 & 86.74 & 86.31 & 86.14 \\
		\expib    & \logisticregression      & \maxweightedavgprobAc & 88.20 & 88.73 & 88.20 & 88.00 & 86.23 & 86.64 & 86.23 & 86.07 \\
		\expib    & \logisticregression      & \summaxprobAc         & 88.02 & 88.65 & 88.02 & 87.84 & 86.57 & 87.08 & 86.57 & 86.45 \\
		\expibdoc & \logisticregression      & none                  & 89.82 & 89.88 & 89.82 & 89.50 & 88.28 & 88.54 & 88.28 & 88.00 \\[1.1mm]
		\expib    & \neuralnetonelayerCPU    & \maxsumprobAc         & 88.37 & 88.72 & 88.37 & 88.06 & 86.14 & 87.24 & 86.14 & 85.36 \\
		\expib    & \neuralnetonelayerCPU    & \maxweightedavgprobAc & 88.20 & 88.64 & 88.20 & 87.90 & 86.40 & 87.43 & 86.40 & 85.68 \\
		\expib    & \neuralnetonelayerCPU    & \summaxprobAc         & 88.45 & 88.82 & 88.45 & 88.17 & 86.48 & 87.56 & 86.48 & 85.82 \\
		\expibdoc & \neuralnetonelayerCPU    & none                  & 89.14 & 89.83 & 89.14 & 88.88 & 89.22 & 89.45 & 89.22 & 88.85 \\[1.1mm]
		\expib    & \randomforest            & \maxsumprobAc         & 71.09 & 78.05 & 71.09 & 69.18 & 70.57 & 78.35 & 70.57 & 68.54 \\
		\expib    & \randomforest            & \maxweightedavgprobAc & 71.34 & 78.21 & 71.34 & 69.44 & 70.66 & 78.78 & 70.66 & 68.63 \\
		\expib    & \randomforest            & \summaxprobAc         & 71.00 & 78.68 & 71.00 & 69.12 & 70.15 & 78.30 & 70.15 & 68.25 \\
		\expibdoc & \randomforest            & none                  & 71.77 & 84.78 & 71.77 & 72.88 & 73.31 & 84.27 & 73.31 & 74.36 \\[1.1mm]
		\expib    & \supportvectorclassifier & \maxsumprobAc         & 87.77 & 88.55 & 87.77 & 87.70 & 85.63 & 87.03 & 85.63 & 85.69 \\
		\expib    & \supportvectorclassifier & \maxweightedavgprobAc & 87.77 & 88.50 & 87.77 & 87.67 & 85.80 & 86.95 & 85.80 & 85.80 \\
		\expib    & \supportvectorclassifier & \summaxprobAc         & 87.51 & 88.33 & 87.51 & 87.42 & 85.20 & 86.67 & 85.20 & 85.27 \\
		\expibdoc & \supportvectorclassifier & none                  & 83.40 & 84.47 & 83.40 & 82.60 & 85.37 & 86.47 & 85.37 & 85.04 \\
		\noalign{\smallskip}\hline\noalign{\smallskip}
	\end{tabular}
\end{table}

\clearpage
\renewcommand*{\refname}{}
\section*{Appendix 2: References}\label{sec:references}
\addcontentsline{toc}{section}{Appendix 2: References}
%
%

\begin{thebibliography}{99.}%
%
\bibitem{tensorflow} M. Abadi et al., \textit{TensorFlow: Large-scale machine learning on heterogeneous
systems}, Software available from tensorflow.org, (2015)
\bibitem{Breiman.1984}L. Breiman, J. Friedman, R.A. Olshen, C.J. Stone, \textit{Classification and Regression Trees}, (Routledge, 1984)
\bibitem{Chawla.2002} N. V. Chawla, K. W. Bowyer, L. O. Hall, W. P. Kegelmeyer, \textit{SMOTE: synthetic minority over-sampling technique}, Journal of artificial intelligence research \textbf{16}, (2002)
\bibitem{Cortes.1995}C. Cortes, V. Vapnik, \textit{Support-vector networks}, Machine Learning \textbf{20}, (1995)
\bibitem{Devlin.2019}J. Devlin, M. Chang, K. Lee, K. Toutanova, \textit{BERT: Pre-training of Deep Bidirectional Transformers for Language Understanding}, (Available via ArXiv, 2018), doi: 10.48550/arXiv.1810.04805
\bibitem{Diez.2023} F. Diez, M. Trebing, S. Schwaar. \textit{Segmentation and aggregation in text classification}, (2023)
\bibitem{Domeniconi.2015} G. Domeniconi, G. Moro, R. Pasolini, C. Sartori, \textit{A Study on Term Weighting for Text Categorization: A Novel Supervised Variant of tf.idf}, International Conference on Data Management Technologies and Applications, (SciTePress, 2015)
\bibitem{Goodfellow.2016}I. Goodfellow, Y. Bengio, A. Courville, \textit{Deep Learning}, (MIT Press, 2016)
\bibitem{Hastie.2009}T. Hastie, R. Tibshirani, J. Friedman, \textit{The Elements of Statistical Learning}, Second Edition: Data Mining, Inference, and Prediction, Springer Series in Statistics, 2nd edn. (Springer New York and Springer Berlin, 2009)
\bibitem{Jurafsky.2008}D. Jurafsky, J. H. Martin, \textit{Speech and language processing: An introduction to natural language processing, computational linguistics, and speech recognition}, (Pearson Prentice Hall, 2008)
\bibitem{Kowsari.2019}K. Kowsari, K. J. Meimandi, M. Heidarysafa, S. Mendu, L. Barnes, D. Brown, \textit{Text Classiﬁcation Algorithms: A Survey}, Information \textbf{10(4)}, (2019)
\bibitem{imblearn} G. Lema{{\^i}}tre, F. Nogueira, C. K. Aridas, \textit{Imbalanced-learn: A Python
Toolbox to Tackle the Curse of Imbalanced Datasets in Machine Learning}, Machine Learning Research \textbf{18}, (2017)
\bibitem{Luhn.1957}H. P. Luhn, \textit{A Statistical Approach to Mechanized Encoding and Searching of Literary Information}, IBM Journal of research and development \textbf{1(4)}, (1957)
\bibitem{MikolovEtAl13}T. Mikolov, K. Chen, G. Corrado, J. Dean, \textit{Efﬁcient estimation of word representations in vector space}, (Available via ArXiv, 2013)
\bibitem{Mockus.1989} J. Mockus, \textit{Bayesian Approach to Global Optimization, Mathematics and its Application}, (Springer Dordrecht, 1989)
\bibitem{sklearn} Pedregosa, F.  et al., \textit{Scikit-learn: Machine Learning in Python}, Journal of Machine Learning Research \textbf{12}, (2011)
\bibitem{PenningtonEtAl14}J. Pennington, R. Socher, C. D. Manning, \textit{GloVe: Global Vectors for Word Representation}. Proceedings of the 2014 Conference on Empirical Methods in Natural Language Processing (EMNLP), (Association for Computational Linguistics, Qatar, 2014), p. 1532-1543
\bibitem{PetersEtAl18}M. E. Peters, M. Neumann, M. Iyyer, M. Gardner, C. Clark, K. Lee, L. Zettlemoyer,
\textit{Deep Contextualized Word Representations}, in: Proceedings of the 2018 Conference of the North American Chapter of the Association for Computational Linguistics: Human Language Technologies, (Association for Computational Linguistics, Louisiana, 2018), p. 2227-2237
\bibitem{Quint.2018}E. Quint, G. Wirka, J. Williams, S. Scott, N. Vinodchandran, \textit{Interpretable classification via supervised variational autoencoders and differentiable decision trees}, (2018)  
\bibitem{Skiena.2017}S. S. Skiena, \textit{The Data science design manual}, (Springer, Cham, 2017)
\bibitem{SPARCKJONES.1972}K. Sparck Jones, \textit{A Statistical Interpretation of Term Speciﬁcity and its Application in Re-
trieval}, Journal of Documentation \textbf{28(1)}, (1972)
\bibitem{Weissweiler}L. Weissweiler, A. Fraser, \textit{Developing a Stemmer for German Based on a Comparative Analysis of Publicly Available Stemmers}, ed. by G. Rehm and T. Declerck. Language Technologies for the Challenges of the Digital Age, p. 81–94
\end{thebibliography}
%

\end{document}